# Selection rules of linear and nonlinear polarization-selective absorption in optically dressed matter

**Michael Feldman*†, Matan Even Tzur†, and Oren Cohen**

Solid State Institute and Physics Department, Technion-Israel Institute of Technology, Haifa 3200003, Israel.

† - These authors contributed equally to this work

**Abstract**

**Dynamical symmetries of laser-dressed matter determine selection rules that determine its absorption spectrum. We explore selection rules for polarization-sensitive absorption in Floquet matter, using Floquet group theory in synthetic dimensions. We present comprehensive tables of selection rules that polarization-structured light impose on Floquet dark states and Floquet dark bands. Notably, our tables encompass nonlinear absorption, to all nonlinear orders, revealing that different nonlinear orders follow distinct polarization selection rules, potentially leading to polarization-tunable optical filters.**

Recently, *Floquet engineering* techniques have been developed for controlling spectral absorption of materials.[1]. For example, in an optically dressed material (described by a time-periodic Hamiltonian [2]) the electronic structure comprises of Floquet states and Floquet bands, and proper engineering of these unique properties can imbue materials with intriguing features. Such Floquet systems are ubiquitous in various areas of physics such as lasers[3], condensed matter[4] and more[4–9].

A Floquet system exhibits spatio-temporal symmetries, also known as dynamical symmetries [10], that determine selection rules, e.g, generation of odd only [10] or circularly polarized [11,12] harmonics in high harmonic generation from isotropic media. Floquet group theory was introduced as a general framework to describe dynamical symmetries, and their associated selection rules on physical observables such as the nonlinear emission spectrum [13]. Notably, it was recently shown (theoretically & experimentally) that DSs impose selection rules also on linear absorption, in the form of symmetry-protected dark states, symmetry-protected dark bands, and symmetry induced transparency [14]. However, nonlinear absorption was not explored. Also, only absorbed light with DS that matches the DS of the absorbing Floquet system was considered. That is, the DSs of the dressing and absorbed fields were matched, so that the full light-matter system exhibited a DS. Notably, it was recently shown that even in the presence of broken dynamical symmetries, selection rules can still be systematically manifested [15,16].

Here we present a generalized framework for absorption and transparency, of an arbitrarily polarized light by dressed matter with DS, even when this probe light does not uphold the DS of the dressed matter. That is, the light-matter system exhibits a broken DS. Nevertheless, the system exhibits dynamical symmetries in space time and synthetic dimensions (SDS), where the synthetic dimensions represent the polarization of the probe light. We derive the selection rules for linear and nonlinear absorption associated with these SDS. We find that polarization-structured dressing light leads to polarization selective Floquet dark states and Floquet dark bands (i.e., polarization-selective absorption). Notably, the derived selection rules encompass nonlinear absorption to all nonlinear orders. We find that different nonlinear orders adhere to different selection rules.

We begin our treatment by considering a specific example that was previously considered - absorption of light by a Benzene molecule, driven (dressed) by a circularly polarized laser field [14]. When the Benzene molecule and driving laser field (with a frequency $\Omega$) are aligned in the XY plane, (Figure 1) the time dependent Hamiltonian of the system is given by[14].

$$1) \qquad \hat{H} = \hat{H}_0 + \lambda \hat{\boldsymbol{d}} \cdot \left( \boldsymbol{Q}_\mathrm{d} e^{i\Omega t} + H.c. \right)$$

in which $\hat{H}_0$ is the time-independent Hamiltonian of the field-free molecule, $\hat{\boldsymbol{d}}$ is dipole moment operator, $\boldsymbol{Q}_\mathrm{d}$ is the complex Jones vector of the pump, and $\lambda$ is proportional to the electric field amplitude of the dressing laser. We note that employing a quantum-optical Hamiltonian[14] is not required, as the semi-classical and quantum-optical Hamiltonians share the same DSs and selection rules. We consider the absorption of a $\hat{z}$ polarized (out of plane) probe beam of frequency $\omega_p$ which interacts with the system (Figure 1.(a)). Including the probe, the interaction term in the Hamiltonian (1)) becomes:

$$2) \qquad \widehat{W} = \lambda \boldsymbol{Q}_\mathrm{d} \cdot \hat{\boldsymbol{d}} e^{i\Omega t} + \lambda \boldsymbol{Q}_\mathrm{p} \cdot \hat{\boldsymbol{d}} \, e^{i\omega_p t} + H.c.$$

in which $\boldsymbol{Q}_\mathrm{d,p}$ are the complex Jones vectors of the complex pump and the probe beams respectively, and $\omega_p$ is the frequency of the prob. Within either linear response theory[14] or Floquet perturbation theory[16], the linear absorption in a frequency $\omega = \omega_p + n\Omega$ is proportional to the susceptibility $I(\omega) \propto -i\tilde{\chi}_n(\omega_p)$ which is given by:

$$\tilde{\chi}_n(\omega_p) = i\lambda^2 \sum_{\nu,\mu,m} \frac{V_{\nu,\mu}^{(-n-m)} V_{\mu,\nu}^{(m)} (p_\nu - p_\mu)}{\epsilon_\mu - \epsilon_\nu + m\Omega - \omega_p - i\gamma_{\nu,\mu}^{(m)}} \quad (3)$$

In which $m$ and $n$ are integers, $\mu, \nu$ are Floquet state indices and $\gamma_{\nu,\mu}^{(m)}$ are dephasing rates. $V_{\mu,\nu}^{(m)} \equiv \frac{1}{\tau}\int_0^\tau \langle u_\mu^{(m)}(t)|\hat{d}|u_\nu^{(m)}(t)\rangle dt$ are transition dipole moments between Floquet states $|u_{\mu,\nu}(t)\rangle$, and $\epsilon_{\mu,\nu}$ are their corresponding quasi-energies, defined by the eigenvalue equation:

$$\left[\hat{H}(t) - \frac{id}{dt}\right]|u_\eta^{(m)}(t)\rangle = \varepsilon_\eta |u_\eta^{(m)}(t)\rangle \quad (4)$$

The dressed system exhibits the DS $\hat{C}_6 = \hat{R}_6 \cdot \hat{\tau}_6$ where $\hat{R}_6$ is spatial rotation of $\frac{2\pi}{6}$ around the $\hat{z}$ axis, and $\hat{\tau}_6$ is $\frac{T}{6} = \frac{2\pi}{6\Omega}$ [13,14] time translation. The DS $\hat{C}_6$ results with symmetry protected dark states (spDS)[14]:

$$\hat{V}_{\nu,\mu}^{(m)} = \begin{cases} 1 & \text{if } e^{i\frac{2\pi}{N}(m_\mu - m_\nu + n)} = 1 \\ 0 & \text{else} \end{cases} \quad m_\mu, m_\nu \in \{0, 1 \ldots N-1\} \quad (5)$$

and symmetry protected Floquet bands:

$$\tilde{\chi}_n(\omega_p) = \begin{cases} 1 & \text{if } e^{i\frac{2\pi}{N}n} = 1 \\ 0 & \text{else} \end{cases} \quad (6)$$

We emphasize that within the linear response approach, Eq. (5-6) are only correct for probes that do not break the $\hat{C}_6$ symmetry, essentially limiting the discussion to $\hat{z}$ polarized probes.

To generalize this result beyond the regime of linear response, and for arbitrarily polarized probe fields, we look for synthetic dynamical symmetries[16] of the perturbed (probed) Hamiltonian $\hat{H} = \hat{H}_0 + \hat{W}$. While the probe may break the $\hat{C}_6$ symmetry imposed by the circularly polarized $\Omega$ field, for any choice of $\boldsymbol{Q}_p$, a reduced dynamical symmetry in synthetic dimensions (synthetic dynamical symmetry – SDS) remains. The SDS is constructed as a composition $\hat{X} = \hat{C}_6 \circ \hat{\zeta}$, where $\hat{\zeta}$ operates in the synthetic space spanned by the polarization vector of the probe $\boldsymbol{Q}_\text{p}$:

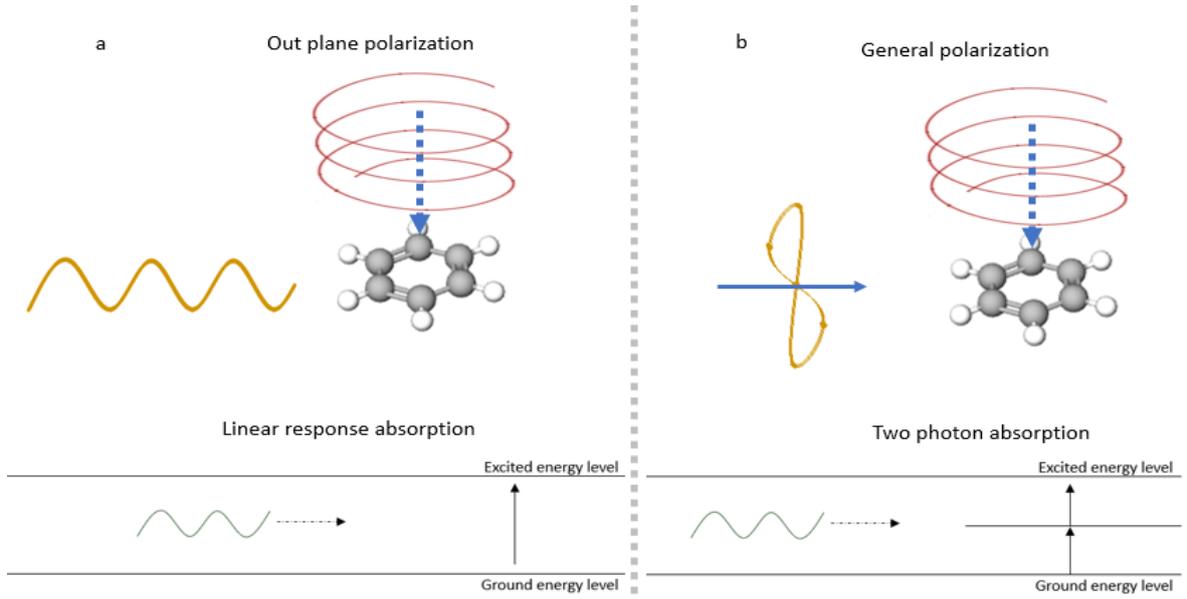

*Figure 1: the different between perturbation theorty (a) and dynamical symmetry (b), on polarizaion of the prob(1) and the absopation order (2)*

Here, $\hat{\zeta}_{\hat{C}_6}$ operates on the synthetic space spanned by the components of the Jones vector $\boldsymbol{Q}_p$, and is given by[16] $\hat{\zeta}_{\hat{C}_6}[\boldsymbol{Q}_p] = e^{\frac{i\omega 2\pi}{6\Omega}}\boldsymbol{Q}_p$. By employing the invariance of the emitted field $\boldsymbol{E}(t)$ (equivalent to the dipole moment expectation value in our semi-classical description) under the SDS operation, we obtain generalized selection rule for symmetry protected dark bands (generalizing Eq.(6)):

$$7) \qquad \boldsymbol{E}(t, \boldsymbol{Q}_p) = \hat{C}_6 \boldsymbol{E}(t, \hat{\zeta}_{\hat{C}_6}[\boldsymbol{Q}_p])$$

This represents the most general selection rule for the absorption of the probe. For example, assuming $\boldsymbol{Q}=q_z\hat{z}$ and that $\boldsymbol{E}$ is linear with $q_z$, Eq. (7)7) reduces to Eq. (6), reproducing the result of ref[14]. However, Eq. (7) contains a wealth of additional information, as it provides selection rules on the nonlinear response to all nonlinear orders, and for arbitrary probe polarization. To obtain it, we denote $\omega \equiv n\Omega + \omega_p$. Fourier expanding Eq. (8):

$$8) \qquad \sum_n \widetilde{\boldsymbol{E}}_\omega(\boldsymbol{Q}_p)e^{i\omega t} = \sum_n \hat{C}_6^{(XY)}\widetilde{\boldsymbol{E}}_\omega\left(\hat{\zeta}_{\hat{C}_6}(\boldsymbol{Q}_p)\right)e^{i\omega t}$$

To first order in $\boldsymbol{Q}$ we obtain

$$9) \quad 1 = e^{\frac{i2\pi}{6}\left(\frac{\omega-\omega_p}{\Omega}\right)} = e^{\frac{i2\pi}{6}n}$$

Which is equivalent to Eq. (6). In a similar way, $\hat{X} = \hat{C}_6 \circ \hat{\zeta}$ imposes a selection rule on the transition between Floquet bands. To derive it, we expand $\langle \nu|\hat{\boldsymbol{d}}|\mu\rangle = \langle \nu|\hat{\boldsymbol{d}}|\mu\rangle \equiv \sum_n \widetilde{\boldsymbol{d}}_n^{(\nu,\mu)}(\boldsymbol{Q}_p)e^{in\omega t}$ in a Fourier series, and employ the relation $\langle \nu|\hat{X}^+ \hat{X}\hat{\boldsymbol{d}} \hat{X}^+\hat{X}|\mu\rangle$. The corresponding generalized selection rule is:

10) $$\widetilde{d}_n^{(\nu,\mu)}(Q_p) = e^{\frac{i2\pi}{6\Omega}(m_\nu - m_\mu + n)} \hat{R}_6 \widetilde{d}_n^{(\nu,\mu)}(\hat{\zeta}_{\hat{C}_6}[Q_p])$$

Eq.(10) is a generalized selection rule for symmetry protected dark-states, which reduces to Eq.(5) for $Q_p \parallel \hat{z}$ and the linear term of $\widetilde{d}_n^{(\nu,\mu)}(Q_p)$. As a generalization, it can also account for the absorption of in-plane probes, e.g., right (RCP) or left (LCP) handed circularly polarized probes with Jones vectors $Q_p \equiv Q(\hat{x} \pm i\hat{y})$ where a plus (minus) corresponds to RCP (LCP) polarized probe. Expanding $\widetilde{d}_n^{(\nu,\mu)}$ into orders of $Q^k$, Eq.(10) becomes

11) $$1 = e^{\frac{i2\pi}{6}\left(n - \left(\frac{\omega_p}{\Omega} \pm 1\right)(k-1)\right)}$$

in which $k$ is the nonlinearity order of the absorption, and $\pm$ corresponds to RCP/LCP, respectively.

Through a similar procedure, we tabulate the generalized selection rules for symmetrical protected bands (Table 1) and symmetry protected dark states (Table 2). The complete derivations are outlined in the SI. For the time-reversal ($\hat{T}$) symmetry, we find that transition between Floquet bands (absorption/ emission) is allowed only for linearly polarized probes that are in phase with the dressing field, for any nonlinearity order. For $\hat{Z}_y$ symmetry, transition between Floquet bands is allowed only for an $\hat{x}$ polarized probe which satisfies $n - \frac{\omega_p}{\Omega}(k-1) + k + 1 = 2l$ and $\hat{y}$ polarized probes satisfying $\frac{\omega_p}{\Omega}(k-1) = 2l$, where $k$ is the nonlinearity order, $l$ is an integer, and $n$ is the Floquet band index. For the circular DS, $\hat{C}_{N,M}$, transition between Floquet bands is allowed by an RCP/LCP probe only satisfying $n - \left(\frac{\omega_p}{\Omega} \pm M\right)(k-1) = Nl$, where a plus (minus) corresponds to RCP (LCP). Similar rules are outlined in Table 2 for symmetry protected dark states.

In summary, this letter derives selection rules for symmetry-protected dark bands and states in Floquet systems by applying the concept of synthetic dynamical symmetries [16]. By employing synthetic dynamical symmetry, we extend the applicability of previous linear response theory treatments to nonlinear absorption by Floquet systems. Notably, our approach enables the identification of selection rules even when one of the fields disrupts the system's DS, accommodating complex field geometry. Our selection rules derived from dynamical symmetry generalized previously reported [14,17] selection rules for absorption in dressed matter. This work holds promise for the development of tunable polarization sensitive optical elements using Floquet dressing. Additionally, it connects the concepts of Floquet engineering and tailored light.

**Table for general SR with other symmetry:**

*Table 1: selection rule of dark bend 'n' on vaires symetry in general order and general polerized. When k is the order of the absorbtion, λ is the magnitude of the light. $\omega_p$ is the frequency of the prob, $\Omega$ is the frequency of the driver, $n, l \in \mathbb{Z}$.*

| Symmetry | Selection rule of $\tilde{\chi}_n^{(k,\hat{x})}, l \in \mathbb{Z}, k(\text{nonlinearity order}) \in \mathbb{N}$ |
|---|---|
| $\hat{T}$ | Transition (absorption/emission): any linearly polarized probe in the same phase with the dressing field for any nonlinearity order. |
| $\hat{Q}$ | Transition (absorption/emission): linearly polarized probe with a phase $\pi k/2$. |
| $\hat{G}$ | Transition (absorption/emission): linearly polarized probe with a phase $\pi\left(n - \frac{\omega_p}{\Omega}(k-1) + k + 1\right)/2$. |
| $\hat{Z}_y$ | $\hat{x}$ polarized probe: $n - \frac{\omega_p}{\Omega}(k-1) + k + 1 = 2l$ |
| | $\hat{y}$ polarized probe: $n - \frac{\omega_p}{\Omega}(k-1) = 2l$ |
| | . |
| $\hat{D}_y$ | $\hat{x}$ polarized probe with a phase $\pi(1+k)/2$ |
| | $\hat{y}$ polarized probe in phase with the dressing field. |
| $\hat{H}_y$ | $\hat{x}$ polarized probe with a phase $\frac{\pi}{2}\left(n - (k-1)\left(\frac{\omega_p}{\Omega}\right) + 1 + k\right)$ |
| | $\hat{y}$ polarized probe with a phase $\frac{\pi}{2}\left(n - \frac{\omega_p}{\Omega}(k-1)\right)$ |
| $\hat{C}_{N,M}$ | RCP polarized probe: $n - \left(\frac{\omega_p}{\Omega} + 1\right)(k-1) = Nl$ |
| | LCP polarized probe: $n - \left(\frac{\omega_p}{\Omega} - 1\right)(k-1) = Nl$ |
| $\hat{e}_{N,M}$ | RCP polarized probe: $n - \left(\frac{\omega_p}{\Omega} + M\right)(k-1) = Nl$ |
| | LCP polarized probe: $n - \left(\frac{\omega_p}{\Omega} - M\right)(k-1) = Nl$ |

Table 2: selection rule betwin two state (μ.ν)at bend 'n' on vaires symetry, in general order and general polerized

| Symmetry | Selection rule of $\hat{V}_{\nu,\mu}^{(n,k)}$ $l \in \mathbb{Z}$, $k(\text{nonlinearity order}) \in \mathbb{N}$ |
|---|---|
| $\hat{T}$ | linearly polarized probe with a phase: $\frac{\pi}{2}(m_\mu - m_\nu)$ |
| $\hat{Q}$ | linearly polarized probe with a phase: $\frac{\pi}{2}(m_\mu - m_\nu + k + 1)$ |
| $\hat{G}$ | linearly polarized probe with a phase: $\frac{\pi}{2}\left(n - \frac{\omega_p}{\Omega}(k-1) + k + 1 + m_\mu - m_\nu\right)$ |
| $\hat{Z}_y$ | $\hat{x}$ polarized probe: $n - \frac{\omega_p}{\Omega}(k-1) + k + 1 + m_\mu - m_\nu = 2l$ |
| | $\hat{y}$ polarized probe: $n - \frac{\omega_p}{\Omega}(k-1) + m_\mu - m_\nu = 2l$ |
| $\hat{D}_y$ | $\hat{x}$ polarized probe with a phase : $\frac{\pi}{2}(1 + k1 + k + m_\mu - m_\nu)$ |
| | $\hat{y}$ polarized probe with a phase : $\frac{\pi}{2}(m_\mu - m_\nu)$ |
| $\hat{H}_y$ | $\hat{x}$ polarized probe with a phase : $\frac{\pi}{2}\left(n - (k-1)\left(\frac{\omega_p}{\Omega}\right) + 1 + k + m_\mu - m_\nu\right)$ |
| | $\hat{y}$ polarized probe with a phase : $\frac{\pi}{2}\left(n - \frac{\omega_p}{\Omega}(k-1) + m_\mu - m_\nu\right)$ |
| $\hat{C}_N$ | RCP polarized probe: $n - \left(\frac{\omega_p}{\Omega} + M\right)(k-1) + m_\mu - m_\nu = Nl$ |
| | LCP polarized probe: $n - \left(\frac{\omega_p}{\Omega} - M\right)(k-1) + m_\mu - m_\nu = Nl$ |
| $\hat{e}_N$ | RCP polarized probe: $n - \left(\frac{\omega_p}{\Omega} + M\right)(k-1) + m_\mu - m_\nu = Nl$ |

LCP polarized probe: $n - \left(\frac{\omega_p}{\Omega} - M\right)(k-1) + m_\mu - m_\nu = Nl$

## Data Availability Statements

The data supporting the findings of this study are available from the corresponding author upon reasonable request.

## Code Availability Statements

The code supporting the findings of this study are available from the corresponding author upon reasonable request.

## Acknowledgments

We thank the Helen Diller Quantum Center for their support. This work was supported by the European Research Council (ERC) under the European Union's Horizon 2020 research and innovation programme (819440-TIMP), and by the Israel Science Foundation (Grant No. 1781/18). M.E.T. gratefully acknowledges the support of the Council for Higher Education scholarship for excellence in quantum science and technology.

## Contributions

All authors make significant contributions to all aspects of the work.

## Competing interests

The authors declare no competing interest.

**Appendix A- derivation**

$$\widehat{W} = \lambda \boldsymbol{Q}_p \cdot \widehat{\boldsymbol{d}}\, e^{i\omega_p t} + H.c.$$

$$\omega = \Omega n + \omega_p$$

$$\boldsymbol{E}_{emit}(\boldsymbol{Q}, t) = \lambda \boldsymbol{Q} \cdot \boldsymbol{r}\, e^{i\omega t}$$

We want to calculate $\tilde{\chi}_n^{(k,\hat{x})}$ for every symmetry $\hat{x}$. First, we find $\hat{\zeta}_{\hat{x}}$ by:

$$\hat{X} = \hat{x} \cdot \hat{\zeta}_{\hat{x}}$$

$$\hat{x} \cdot \hat{\zeta}_{\hat{x}}(\widehat{W}) = \widehat{W}$$

second, we find the SR of the emitter field by:

$$\hat{x} \cdot \hat{\zeta}_{\hat{T}}\, \boldsymbol{E}(t, \boldsymbol{Q}) = \boldsymbol{E}(t, \boldsymbol{Q})$$

Which $\boldsymbol{E}(t, \boldsymbol{Q})$ is the expectation value of the field. Additionally, we use Fourier series to separate the frequencies, and Tailor series to separate the order of the absorption.

**Derivation of T-symmetry:**

The symmetry operator:

$$\hat{X} = \hat{T} \cdot \hat{\zeta}_{\hat{T}}$$

$$\hat{T} \cdot \hat{\zeta}_{\hat{T}} \lambda \boldsymbol{Q}_p \cdot \boldsymbol{r}\, e^{i\omega_p t} + h.c = \hat{\zeta}_{\hat{T}} \lambda \boldsymbol{Q}_p \cdot \boldsymbol{r}\, e^{-i\omega_p t} + h.c$$

Therefore, the synthetic operator is:

$$\hat{\zeta}_{\hat{T}}(\lambda \boldsymbol{Q}_p) = \lambda \overline{\boldsymbol{Q}_p}$$

Now we looked at the expectation value:

$$\hat{T} \cdot \hat{\zeta}_{\hat{T}}\, \boldsymbol{E}(t, \boldsymbol{Q}) = \boldsymbol{E}(t, \boldsymbol{Q})$$

$$\hat{T} \cdot \hat{\zeta}_{\hat{T}}\, \boldsymbol{E}(t, \boldsymbol{Q}) = \sum_n \boldsymbol{E}_n(\overline{\boldsymbol{Q}}) e^{-i\omega t} = \sum_n \boldsymbol{E}_n(\boldsymbol{Q}) e^{i\omega t}$$

$$\boldsymbol{E}_n(\overline{\boldsymbol{Q}}) = \boldsymbol{E}_{-n}(\boldsymbol{Q}) = \overline{\boldsymbol{E}}_n(\boldsymbol{Q})$$

Therefor regardless to the polarization the SR is:

$$\boldsymbol{E}_{k,n} \in \mathbb{R}$$

**Derivation of Q-symmetry:**

$$\hat{X} = \hat{Q} \cdot \hat{\zeta}_{\hat{Q}}$$

For $\hat{x}$ or $\hat{y}$ polarized:

$$\hat{Q} \cdot \hat{\zeta}_{\hat{Q}} (\lambda \hat{x}\, e^{i\omega_p t} + h.c) = -\hat{\zeta}_{\hat{Q}} (\hat{x}\, e^{-i\omega_p t} + h.c)$$

Therefore, the synthetic operator is:

$$\hat{\zeta}_{\hat{Q}}(\lambda \boldsymbol{Q}_p) = -\lambda \overline{\boldsymbol{Q}_p}$$

Now we looked at expectation value, we write for $\hat{x}$ polarized, but is the same for $\hat{y}$ polarized:

$$\hat{Q} \cdot \hat{\zeta}_{\hat{Q}}\, \boldsymbol{E}(t,\boldsymbol{Q}) = \boldsymbol{E}(t,\boldsymbol{Q})$$

$$\hat{Q} \cdot \hat{\zeta}_{\hat{Q}}\, \boldsymbol{E}(t,\boldsymbol{Q}) = \sum_n \boldsymbol{E}_n\left(\hat{\zeta}_{\hat{Q}}(\boldsymbol{Q})\right) e^{-i\omega t} = \sum_n -\boldsymbol{E}_n(-\overline{\boldsymbol{Q}}) e^{-i\omega t} = \sum_n \boldsymbol{E}_n(\boldsymbol{Q}) e^{i\omega t}$$

$$\overline{\boldsymbol{E}}_n(Q) = -\boldsymbol{E}_n(-\overline{Q})$$

$$\sum_k \overline{\boldsymbol{E}}_{n,k} \lambda^k = \sum_k -\boldsymbol{E}_{n,k}(-\lambda)^k$$

$$\overline{\boldsymbol{E}}_{n,k} = (-1)^{k+1} \boldsymbol{E}_{n,k}$$

the SR is:

$$i^{1+k} \cdot \boldsymbol{E}_{k,n} \in \mathbb{R}$$

**Derivation of G-symmetry:**

$$\hat{X} = \hat{G} \cdot \hat{\zeta}_{\hat{G}}$$

For $\hat{x}$ or $\hat{y}$ polarized:

$$\hat{G} \cdot \hat{\zeta}_{\hat{G}} (\lambda \hat{x}\, e^{i\omega_p t} + h.c) = -\hat{\zeta}_{\hat{G}} \left(\hat{x}\, e^{-i\omega_p t}(-1)^{\frac{\omega_p}{\Omega}} + h.c\right)$$

Therefore, the synthetic operator is:

$$\hat{\zeta}_{\hat{G}}(\lambda \mathbf{Q}_{\mathrm{p}}) = (-1)^{-\frac{\omega_p}{\Omega}-1}\lambda\overline{\mathbf{Q}_{\mathrm{p}}}$$

Now we looked at the expectation value, we write for $\hat{x}$ polarized, but is the same for $\hat{y}$ polarized:

$$\hat{G} \cdot \hat{\zeta}_{\hat{G}}\, \mathbf{E}(t,\mathbf{Q}) = \mathbf{E}(t,\mathbf{Q})$$

$$\hat{G} \cdot \hat{\zeta}_{\hat{G}}\, \mathbf{E}(t,\mathbf{Q}) = \sum_n (-1)^{\frac{\omega}{\Omega}+1} \mathbf{E}_n\!\left(\hat{\zeta}_{\hat{G}}(\mathbf{Q})\right) e^{-i\omega t} = \sum_n \mathbf{E}_n\!\left(\overline{\mathbf{Q}}(-1)^{-\frac{\omega_p}{\Omega}-1}\right) e^{-i\omega t}$$

$$= \sum_n \mathbf{E}_n(\mathbf{Q}) e^{i\omega t}$$

$$\overline{\mathbf{E}}_n(\mathbf{Q}) = (-1)^{\frac{\omega}{\Omega}+1} \mathbf{E}_n\!\left(\overline{\mathbf{Q}}(-1)^{-\frac{\omega_p}{\Omega}-1}\right)$$

$$\sum_k \overline{\mathbf{E}}_{n,k} \lambda^k = \sum_k (-1)^{\frac{\omega}{\Omega}+1} \mathbf{E}_{n,k}((-1)^{-\frac{\omega_p}{\Omega}-1}\lambda)^k$$

$$\overline{\mathbf{E}}_{n,k} = (-1)^{\frac{\omega}{\Omega}+1-\left(\frac{\omega_p}{\Omega}+1\right)k} \mathbf{E}_{n,k} = (-1)^{\frac{\omega+\omega_p}{\Omega}+1-1-\left(\frac{\omega_p}{\Omega}+1\right)(k-1)} \mathbf{E}_{n,k}$$

The SR is:

$$i^{n-\left(\frac{\omega_p}{\Omega}+1\right)(k-1)} \cdot \mathbf{E}_{n,k} \in \mathbb{R}$$

**Derivation of $Z_y$-symmetry:**

$$\hat{X} = \hat{Z} \cdot \hat{\zeta}_{\hat{Z}}$$

For $\hat{x}$ polarized:

$$\hat{Z} \cdot \hat{\zeta}_{\hat{Z}}(\lambda \hat{x}\, e^{i\omega_p t} + h.c) = -\hat{\zeta}_{\hat{Z}}\!\left(\hat{x}\, e^{i\omega_p t}(-1)^{\frac{\omega_p}{\Omega}} + h.c\right)$$

Therefore, the synthetic operator is:

$$\hat{\zeta}_{\hat{Z}}(\lambda \mathbf{Q}_{\mathrm{p}}) = (-1)^{-\frac{\omega_p}{\Omega}-1} \lambda \mathbf{Q}_{\mathrm{p}}$$

Now we looked at the expectation value of the emitter field:

$$\hat{Z}\cdot\hat{\zeta}_{\hat{Z}}\,\boldsymbol{E}(t,\boldsymbol{Q}) = \sum_n (-1)^{\frac{\omega}{\Omega}+1}\boldsymbol{E}_n\left(\hat{\zeta}_{\hat{Z}}(\boldsymbol{Q})\right)e^{i\omega t} = \sum_n \boldsymbol{E}_n\left((-1)^{-\frac{\omega_p}{\Omega}-1}\boldsymbol{Q}\right)e^{i\omega t}$$

$$= \sum_n \boldsymbol{E}_n(\boldsymbol{Q})e^{i\omega t}$$

$$\boldsymbol{E}_n(Q) = (-1)^{\frac{\omega}{\Omega}+1}\boldsymbol{E}_n\left(\boldsymbol{Q}(-1)^{-\frac{\omega_p}{\Omega}-1}\right)$$

$$\sum_k \boldsymbol{E}_{n,k}\lambda^k = \sum_k (-1)^{\frac{\omega}{\Omega}+1}\boldsymbol{E}_{n,k}((-1)^{-\frac{\omega_p}{\Omega}-1}\lambda)^k$$

$$\boldsymbol{E}_{n,k} = (-1)^{\frac{\omega}{\Omega}+1-\left(\frac{\omega_p}{\Omega}+1\right)k}\boldsymbol{E}_{n,k} = (-1)^{\frac{\omega+\omega_p}{\Omega}-\left(\frac{\omega_p}{\Omega}+1\right)(k-1)}\boldsymbol{E}_{n,k}$$

The SR is:

$$n - \left(\frac{\omega_p}{\Omega}+1\right)(k-1) = 2l\,,\,l\in\mathbb{Z}$$

For $\hat{y}$ polarized:

$$\hat{Z}\cdot\hat{\zeta}_{\hat{Z}}\left(\lambda\hat{y}\,e^{i\omega_p t}+h.c\right) = \hat{\zeta}_{\hat{Z}}\left(\hat{x}\,e^{i\omega_p t}(-1)^{\frac{\omega_p}{\Omega}}+h.c\right)$$

Therefore, the synthetic operator is:

$$\hat{\zeta}_Z(\lambda\boldsymbol{Q}_{\mathrm{p}}) = (-1)^{-\frac{\omega_p}{\Omega}}\lambda\boldsymbol{Q}_{\mathrm{p}}$$

Now we looked at the expectation value of the emitter field:

$$\hat{Z}\cdot\hat{\zeta}_{\hat{Z}}\,\boldsymbol{E}(t,\boldsymbol{Q}) = \sum_n (-1)^{\frac{\omega}{\Omega}}\boldsymbol{E}_n\left(\hat{\zeta}_{\hat{Z}}(\boldsymbol{Q})\right)e^{i\omega t} = \sum_n (-1)^{\frac{\omega}{\Omega}}\boldsymbol{E}_n\left((-1)^{-\frac{\omega_p}{\Omega}}\boldsymbol{Q}\right)e^{i\omega t}$$

$$= \sum_n \boldsymbol{E}_n(\boldsymbol{Q})e^{i\omega t}$$

$$\boldsymbol{E}_n(Q) = (-1)^{\frac{\omega}{\Omega}}\boldsymbol{E}_n\left(\boldsymbol{Q}(-1)^{-\frac{\omega_p}{\Omega}}\right)$$

$$\sum_k \boldsymbol{E}_{n,k}\lambda^k = \sum_k (-1)^n\boldsymbol{E}_{n,k}((-1)^{-\frac{\omega_p}{\Omega}}\lambda)^k$$

$$\boldsymbol{E}_{n,k} = (-1)^{\frac{\omega}{\Omega}-\left(\frac{\omega_p}{\Omega}\right)k}\boldsymbol{E}_{n,k} = (-1)^{\frac{\omega+\omega_p}{\Omega}-\left(\frac{\omega_p}{\Omega}\right)(k-1)}\boldsymbol{E}_{n,k}$$

The SR is:

$$n - \left(\frac{\omega_p}{\Omega}\right)(k-1) = 2l\, ,\, l \in \mathbb{Z}$$

**Derivation of D-symmetry:**

The symmetry operator:

$$\hat{X} = \hat{D} \cdot \hat{\zeta}_{\hat{D}}$$

For $\hat{x}$ polarized:

$$\hat{D} \cdot \hat{\zeta}_{\hat{D}}\left(\lambda \hat{x}\, e^{i\omega_p t} + h.c\right) = \hat{\zeta}_{\hat{D}}\left(-\hat{x}\, e^{-i\omega_p t} + h.c\right)$$

Therefore, the synthetic operator is:

$$\hat{\zeta}_{\hat{D}}(\lambda \boldsymbol{Q}_p) = -\lambda \overline{\boldsymbol{Q}}_p$$

From here is the same as $\hat{Q}$ symmetry therefore The SR is:

$$i^{1+k} \cdot \boldsymbol{E}_{k,n} \in \mathbb{R}$$

And for $\hat{y}$ polarized:

$$\hat{D} \cdot \hat{\zeta}_{\hat{D}}\left(\lambda \hat{y}\, e^{i\omega_p t} + h.c\right) = \hat{\zeta}_{\hat{D}}\left(\hat{y}\, e^{-i\omega_p t} + h.c\right)$$

Therefore, the synthetic operator is:

$$\hat{\zeta}_{\hat{D}}(\lambda \boldsymbol{Q}_p) = \lambda \overline{\boldsymbol{Q}}_p$$

And we get operator like a $\hat{T}$ symmetry, therefore The SR is:.

$$\boldsymbol{E}_{k,n} \in \mathbb{R}$$

**Derivation of H-symmetry:**

The symmetry operator:

$$\hat{X} = \hat{H}_y \cdot \hat{\zeta}_{\hat{H}}$$

For $\hat{x}$ polarized:

$$\hat{H} \cdot \hat{\zeta}_{\hat{H}}\left(\lambda \hat{x}\, e^{i\omega_p t} + h.c\right) = \hat{\zeta}_{\hat{H}}\left(-\hat{x}\, e^{-i\omega_p t} + h.c\right)$$

Therefore, the synthetic operator is:

$$\hat{\zeta}_{\hat{H}}(\lambda \boldsymbol{Q}_\mathrm{p}) = -\lambda \overline{\boldsymbol{Q}}_\mathrm{p}(-1)^{-\frac{\omega_p}{\Omega}}$$

From here is the same as $\hat{G}$ symmetry therefore the selection rule is:

$$i^{n-\left(\frac{\omega_p}{\Omega}+1\right)(k-1)} \cdot \boldsymbol{E}_{n,k} \in \mathbb{R}$$

For $\hat{y}$ polarized:

$$\hat{H} \cdot \hat{\zeta}_{\hat{H}}(\lambda \hat{y} \, e^{i\omega_p t} + h.c) = \hat{\zeta}_{\hat{H}}(\hat{y} \, e^{-i\omega_p t} + h.c)$$

Therefore, the synthetic operator is:

$$\hat{\zeta}_{\hat{H}}(\lambda \boldsymbol{Q}_\mathrm{p}) = \lambda \overline{\boldsymbol{Q}}_\mathrm{p}(-1)^{-\frac{\omega_p}{\Omega}}$$

Now we looked at the expectation value of the emitter field:

$$\hat{H} \cdot \hat{\zeta}_{\hat{H}} \, \boldsymbol{E}(t, \boldsymbol{Q}) = \sum_n (-1)^{\frac{\omega}{\Omega}} \boldsymbol{E}_n\left(\hat{\zeta}_{\hat{H}}(\boldsymbol{Q})\right) e^{-i\omega t} = \sum_n (-1)^{\frac{\omega}{\Omega}} \boldsymbol{E}_n\left((-1)^{-\frac{\omega_p}{\Omega}} \overline{\boldsymbol{Q}}\right) e^{-i\omega t}$$

$$= \sum_n \boldsymbol{E}_n(\boldsymbol{Q}) e^{i\omega t}$$

$$(-1)^{\frac{\omega}{\Omega}} \boldsymbol{E}_n\left((-1)^{-\frac{\omega_p}{\Omega}} \overline{\boldsymbol{Q}}\right) = \overline{\boldsymbol{E}}_n(\boldsymbol{Q}) e^{i\omega t}$$

$$\sum_k \overline{\boldsymbol{E}}_{n,k} \lambda^k = \sum_k (-1)^{\frac{\omega}{\Omega}} \boldsymbol{E}_{n,k}((-1)^{-\frac{\omega_p}{\Omega}} \lambda)^k$$

$$\overline{\boldsymbol{E}}_{n,k} = (-1)^{\frac{\omega}{\Omega} - \frac{\omega_p}{\Omega} k} \boldsymbol{E}_{n,k} = (-1)^{n - \frac{\omega_p}{\Omega}(k-1)} \boldsymbol{E}_{n,k}$$

The SR is:

$$i^{n - \frac{\omega_p}{\Omega}(k-1)} \boldsymbol{E}_{n,k} \in \mathbb{R}$$

**Derivation of C-symmetry:**

The symmetry operator:

$$\hat{X} = \hat{C}_{N,M} \cdot \hat{\zeta}_{\hat{C}}$$

The vector $\hat{e}_{RH}, \hat{e}_{LH}$ are RH, LH polarizations: $x+iy, x-iy$. these vectors are the eigen vectors of the rotation operators, with eigen value of $e^{\pm \frac{i2\pi}{N}M}$.

For RCP polarized:

$$\hat{C} \cdot \hat{\zeta}_{\hat{C}}(\lambda \hat{e}_{RH} e^{i\omega_p t} + h.c) = \hat{\zeta}_{\hat{C}} \left( e^{\frac{i2\pi}{N}M} \lambda \hat{e}_{RH} e^{-i\omega_p t} e^{2\pi \frac{i\omega_p}{N\Omega}} + h.c \right)$$

Therefore, the synthetic symmetry operator is:

$$\hat{\zeta}_{\hat{C}}(\lambda \mathbf{Q}_p) = \lambda \mathbf{Q}_p e^{-2\pi \frac{i\omega_p}{N\Omega}} e^{-\frac{i2\pi}{N}M}$$

Now we looked at the expectation value of the emitter field:

$$\hat{C} \cdot \hat{\zeta}_{\hat{C}} \, \mathbf{E}(t,\mathbf{Q}) = \sum_n e^{\frac{i2\pi}{N}M} e^{2\pi i \frac{\omega}{N\Omega}} \mathbf{E}_n \left( \hat{\zeta}_{\hat{C}}(\mathbf{Q}) \right) e^{i\omega t}$$

$$= \sum_n e^{\frac{i2\pi}{N}M} e^{2\pi \frac{i\omega}{N\Omega}} \mathbf{E}_n \left( e^{-\frac{i2\pi}{N}M} e^{-2\pi i \frac{\omega_p}{N\Omega}} \mathbf{Q} \right) e^{i\omega t} = \sum_n \mathbf{E}_n(\mathbf{Q}) e^{i\omega t}$$

$$e^{\frac{i2\pi}{N}M} e^{2\pi i \frac{\omega}{N\Omega}} \mathbf{E}_n \left( e^{-\frac{i2\pi}{N}M} e^{-2\pi i \frac{\omega_p}{N\Omega}} \mathbf{Q} \right) = \mathbf{E}_n(\mathbf{Q}) e^{i\omega t}$$

$$\sum_k \mathbf{E}_{n,k} \lambda^k = \sum_k e^{\frac{i2\pi}{N}M} e^{2\pi i \frac{\omega}{N\Omega}} \mathbf{E}_{n,k} (e^{-\frac{i2\pi}{N}M} e^{-2\pi i \frac{\omega_p}{N\Omega}} \lambda)^k$$

$$\mathbf{E}_{n,k} = \mathbf{E}_{n,k} e^{\frac{2\pi i}{N}\left(\frac{\omega}{\Omega}+M-\left(\frac{\omega_p}{\Omega}+M\right)k\right)} = \mathbf{E}_{n,k} e^{\frac{2\pi i}{6}\left(\frac{\omega-\omega_p}{\Omega}-\left(\frac{\omega_p}{\Omega}+M\right)(k-1)\right)}$$

$$n + \left(\frac{\omega_p}{\Omega} + M\right)(k-1) = Nl, l \in \mathbb{Z}$$

For LCP polarized:

$$\hat{C} \cdot \hat{\zeta}_{\hat{C}}(\lambda \hat{e}_{LH} e^{i\omega_p t} + h.c) = \hat{\zeta}_{\hat{C}} \left( e^{-\frac{i2\pi}{N}M} \lambda \hat{e}_{RH} e^{-i\omega_p t} e^{2\pi \frac{i\omega_p}{N\Omega}} + h.c \right)$$

Therefore, the synthetic symmetry operator is:

$$\hat{\zeta}_{\hat{C}}(\lambda \mathbf{Q}_p) = \lambda \mathbf{Q}_p e^{-2\pi \frac{i\omega_p}{N\Omega}} e^{\frac{i2\pi}{N}M}$$

Now we looked at the expectation value of the emitter field:

$$\hat{C} \cdot \hat{\zeta}_{\hat{C}} \, \mathbf{E}(t,\mathbf{Q}) = \sum_n e^{-\frac{i2\pi}{N}M} e^{2\pi i \frac{\omega}{N\Omega}} \mathbf{E}_n \left( \hat{\zeta}_{\hat{Z}}(\mathbf{Q}) \right) e^{i\omega t}$$

$$= \sum_n e^{-\frac{i2\pi}{N}M} e^{2\pi \frac{i\omega}{N\Omega}} \mathbf{E}_n \left( e^{\frac{i2\pi}{N}M} e^{-2\pi i \frac{\omega_p}{N\Omega}} \mathbf{Q} \right) e^{i\omega t} = \sum_n \mathbf{E}_n(\mathbf{Q}) e^{i\omega t}$$

$$e^{-\frac{i2\pi}{N}M}e^{2\pi i\frac{\omega}{N\Omega}}E_n\left(e^{\frac{i2\pi}{N}M}e^{-2\pi i\frac{\omega_p}{N\Omega}}Q\right)=E_n(Q)e^{i\omega t}$$

$$\sum_k E_{n,k}\lambda^k = \sum_k e^{-\frac{i2\pi}{N}M}e^{2\pi i\frac{\omega}{N\Omega}}E_{n,k}(e^{-\frac{i2\pi}{N}M}e^{-2\pi i\frac{\omega_p}{N\Omega}}\lambda)^k$$

$$E_{n,k}=E_{n,k}e^{\frac{2\pi i}{N}\left(\frac{\omega}{\Omega}-M-\left(\frac{\omega_p}{\Omega}-M\right)k\right)}=E_{n,k}e^{\frac{2\pi i}{N}\left(\frac{\omega-\omega_p}{\Omega}-\left(\frac{\omega_p}{\Omega}-M\right)(k-1)\right)}$$

The selection rule is:

$$n-\left(\frac{\omega_p}{\Omega}-M\right)(k-1)=Nl\,,l\in\mathbb{Z}$$

**Derivation of e-symmetry:**

The symmetry operator:

$$\hat{X}=\hat{e}_{N,M}\cdot\hat{\zeta}_{\hat{C}}$$

Now we define $\hat{e}_{RH/LH}$, as $x\pm iby$ the right\left elliptical polarized. Thise units vectors are the eigen vector of the operator: $\hat{L}_b\hat{R}_N\hat{L}_{\frac{1}{b}}$ and they have eigen value of $e^{\pm\frac{i2\pi}{N}M}$

For RCP polarized:

$$\hat{e}\cdot\hat{\zeta}_{\hat{e}}(\lambda\hat{e}_{RH}\,e^{i\omega_p t}+h.c)=\hat{\zeta}_{\hat{e}}\left(e^{\frac{i2\pi}{N}M}\lambda\hat{e}_{RH}e^{-i\omega_p t}e^{2\pi\frac{i\omega_p}{N\Omega}}+h.c\right)$$

Therefore, the synthetic symmetry operator is:

$$\hat{\zeta}_{\hat{e}}(\lambda Q_p)=\lambda Q_p e^{-2\pi\frac{i\omega_p}{N\Omega}}e^{-\frac{i2\pi}{N}M}$$

From here is the same as $\hat{C}$ symmetry therefore the selection rule is:

$$n-\left(\frac{\omega_p}{\Omega}+M\right)(k-1)=Nl\,,l\in\mathbb{Z}$$

For LCP polarized:

$$\hat{e}\cdot\hat{\zeta}_{\hat{e}}(\lambda\hat{e}_{RH}\,e^{i\omega_p t}+h.c)=\hat{\zeta}_{\hat{e}}\left(e^{\frac{i2\pi}{N}M}\lambda\hat{e}_{RH}e^{-i\omega_p t}e^{2\pi\frac{i\omega_p}{N\Omega}}+h.c\right)$$

Therefore, the synthetic symmetry operator is:

$$\hat{\zeta}_{\hat{e}}(\lambda Q_p)=\lambda Q_p e^{-2\pi\frac{i\omega_p}{N\Omega}}e^{-\frac{i2\pi}{N}M}$$

From here is the same as $\hat{C}$ symmetry therefore the selection rule is:

$$n - \left(\frac{\omega_p}{\Omega} - M\right)(k-1) = 6l, l \in \mathbb{Z}$$

The derivation of $\hat{V}_{\nu,\mu}^{(n,k)}$ is very similar to the derivation of $\tilde{\chi}_n^{(k,\hat{x})}$, but now the expectation value of $\boldsymbol{E}_{emit}$ added the eigen value of the state $\mu, \nu$. Therefore, we need to solve:

$$e^{\frac{i2\pi}{N}(m_\mu - m_\nu)}\hat{x} \cdot \hat{\zeta}_{\hat{T}} \, \boldsymbol{E}(t, \mathbf{Q}) = \boldsymbol{E}(t, \mathbf{Q})$$

Which N is the number of the different state in every band. Because of that we get the same SR but with addition if $m_\mu - m_\nu$.